# Ferromagnetism in the Hubbard model: instability of the Nagaoka state on the triangular, honeycomb and kagome lattices *

Thoralf Hanisch, Burkhard Kleine, Afra Ritzl
and Erwin Müller-Hartmann **
Institut für Theoretische Physik
Universität zu Köln
Zülpicher Straße 77, D - 50937 Köln
Federal Republic of Germany

January 24, 1995

cond-mat/9501116  24 Jan 1995

## Abstract

In order to analyse the lattice dependence of ferromagnetism in the two-dimensional Hubbard model we investigate the instability of the fully polarized ferromagnetic ground state (Nagaoka state) on the triangular, honeycomb and kagome lattices. We mainly focus on the local instability, applying single spin flip variational wave functions which include majority spin correlation effects. The question of global instability and phase separation is addressed in the framework of Hartree-Fock theory. We find a strong tendency towards Nagaoka ferromagnetism on the non-bipartite lattices (triangular, kagome) for more than half filling. For the triangular lattice we find the Nagaoka state to be unstable above a critical density of $n = 1.887$ at $U = \infty$, thereby significantly improving former variational results. For the kagome lattice the region where ferromagnetism prevails in the phase diagram widely exceeds the flat band regime. Our results even allow the stability of the Nagaoka state in a small region below half filling. In the case of the bipartite honeycomb lattice several disconnected regions are left for a possible Nagaoka ground state.

# 1  Introduction

The Hubbard model is well established today as the simplest model to describe strongly correlated electrons on a lattice. It contains hopping processes for each spin direction (one orbital per lattice site) and an on-site repulsion as a simple approximation to the Coulomb interaction between the electrons. One of the most important motivations for introducing the Hubbard model thirty years ago [1, 2, 3] was to obtain a better understanding of the appearance of ferromagnetism in transition metals. However, the investigation of the existence of ferromagnetic phases in the Hubbard model, in particular the proof of the stability of the fully polarized ferromagnetic state, the so-called Nagaoka state, turned out to be a very hard task. Up to now only very few exact results are known about the existence of a Nagaoka ground state in the Hubbard model. Most of them are extensions of the famous Nagaoka theorem [4] and deal with the very special case of a half-filled lattice plus one single electron at infinite on-site repulsion $U$ (for example [5, 6]). In this situation the Nagaoka state is found to be stable on most of the common lattices like square lattice, *sc*-, *fcc*-, *bcc*-lattice or triangular lattice. For bipartite lattices this stability holds also for the particle-hole symmetric case of one hole in an otherwise half-filled band.

The first thermodynamically relevant proof of the stability of the Nagaoka state was presented by Mielke and Tasaki [7, 8, 9] for a special class of two-dimensional lattices with flat energy bands, for example the kagome lattice. For all other lattices progress was only achieved in restricting the region where the Nagaoka state might be a ground state. Starting with the work of Shastry, Krishnamurthy and Anderson (SKA) [10] the instability of the Nagaoka state was investigated using single spin flip variational wave functions [11, 12, 13]. This work focussed mainly on the square lattice. Two of the present authors achieved an improvement of former results by including majority spin correlation effects into the variational ansatz [14], again for the square lattice.

As demonstrated by Mattis [15] two criteria are essential for the Nagaoka mechanism to produce ferromagnetism: there should be loops in the underlying lattice (which is not fulfilled for the one-dimensional chain e.g.) and there should be constructive interference of different paths of particles or holes around loops. Like other non-bipartite lattices the triangular lattice shows the shortest possible loop length three and is therefore a good candidate for Nagaoka ferromagnetism in the case of more than half filling. For less than half filling, however, the Nagaoka state is expected to be suppressed because the second criterion is violated.

In the first part of the present paper we use various methods to investigate the instability of the Nagaoka state on the triangular lattice. In order to obtain a general view of the phase diagram we start with the Hartree-Fock approximation (section 2.1). In this context we investigate ferromagnetic states as well as other spin structures like the 120° configuration with three sublattices. Allowing the possibility of phase separation we achieve a large reduction of the Nagaoka stability region near half filling. In section 2.2 a systematic analysis of the stability of the Nagaoka state with respect to a single spin flip is performed by applying some of the extended variational wave functions of [14] to the triangular lattice. This requires an efficient method for calculating the elements of the one particle density matrix, which we describe in detail in the Appendix. We are able to show that the Nagaoka state is not stable for all densities above half filling at $U = \infty$. For one hole less than half filling the Nagaoka theorem is known to be not valid for non-bipartite lattices. We study this case in more detail in section 2.3 and demonstrate that a global Nagaoka instability follows from rather general considerations which apply not only to the triangular lattice.

In sections 3 and 4 we continue our study of Nagaoka ferromagnetism on two-dimensional lattices with the honeycomb and the kagome lattice. The densities of states of both lattices are closely related to the DOS of the triangular lattice. Again we use variational single spin flip wave functions to establish regions of guaranteed instability of the Nagaoka state in the phase diagram. For the bipartite honeycomb lattice two types of regions are left for a possible Nagaoka ground state in which different physical mechanisms apply. The non-bipartite kagome lattice has been the subject of growing attention in recent years. The interest arises primarily because a two-dimensional kagome like structure is found in real substances [16]. Furthermore, the existence of a flat energy band in the nearest neighbour hopping model gives rise to interesting physics. For example, it implies that the Nagaoka state is stable in the flat band region since a ferromagnetic alignment of the spins requires no additional cost of kinetic energy and guarantees the absence of double occupancies. However, as Mielke [7] pointed out there is a degeneracy of the ground state, including states with all possible values of the total spin. Nevertheless, using statistical arguments he showed that for a certain density range states with extensive total spin dominate independently of the strength of the Coulomb interaction $U$. In our work we go beyond the region considered by Mielke deriving variational criteria for the instability of the Nagaoka state in the whole range $0 < n < 2$ and for all possible values of $U$.

# 2 Triangular lattice

By diagonalizing the hopping part of the conventional single band Hubbard model [17]

$$\mathcal{H} = -t \sum_{\langle i,j \rangle, \sigma} c_{i\sigma}^+ c_{j\sigma} + U \sum_i n_{i\uparrow} n_{i\downarrow} \tag{1}$$

on the triangular lattice we obtain the band dispersion

$$\varepsilon_\triangle(\mathbf{k}) = -t \left( 2\cos(k_x) + 4\cos\left(\frac{k_x}{2}\right) \cos\left(\frac{\sqrt{3}}{2} k_y\right) \right) \tag{2}$$

where $\mathbf{k}$ belongs to the hexagon-shaped first Brillouin zone (BZ). For $t > 0$ the lower band edge $\varepsilon_b = -z_\triangle t = -6t$ is found at $\mathbf{k} = \mathbf{0}$, whereas the upper band edge with the energy $\varepsilon_t = 3t$ is reached at all the corners of the hexagon. The density of states $\rho_\triangle(\varepsilon) = \langle \delta(\varepsilon - \varepsilon_\triangle(\mathbf{k})) \rangle_{BZ}$ (expressed by a complete elliptic integral in the Appendix) is depicted in Fig. 1. It shows a logarithmic van Hove singularity at $\varepsilon = 2t$ where the Fermi surface forms a *hexagon* with a volume of exactly 3/4 of the whole BZ and contains the saddle points of the band (2) (e.g. $\mathbf{Q} = (0, 2\pi/\sqrt{3})$ ) as corners.

In the phase diagrams displayed in this paper we will always represent the on-site repulsion $U$ in terms of $U_{red} = U/(U + U_{BR})$ where $U_{BR}$ denotes the Brinkman-Rice critical coupling [18]. For the triangular lattice $U_{BR} = 15.81 |t|$.

## 2.1 Phase diagram in HF-approximation

In this subsection we present a phase diagram of the triangular lattice in HF-approximation, thereby extending previous studies of [19, 20, 21]. In particular, we obtain a first instability region for the Nagaoka state which will be extended in the following subsections.

Let us first briefly review HF-theory: In (unrestricted) HF-theory the two particle interaction is replaced by

$$\sum_i U n_{i\uparrow} n_{i\downarrow} \rightarrow \sum_i \frac{U}{2} \left( \langle \delta n_i \rangle n_i - 4\langle \mathbf{S}_i \rangle \mathbf{S}_i - \frac{1}{2}( \langle \delta n_i \rangle^2 - 4\langle \mathbf{S}_i \rangle^2 - n^2) \right),$$

where $n$ stands for the averaged particle density, $\delta n_i := n_i - n$ denotes the charge density fluctuations and $\mathbf{S}_i$ corresponds to the spin operators at site $i$.

This ansatz simplifies if one restricts the theory to a uniform charge density ($\langle \delta n_i \rangle = 0$) and to a helical spin density wave $\langle \mathbf{S}_i \rangle = \frac{m}{2} (\cos(\mathbf{QR}_i)\mathbf{e}_x + \sin(\mathbf{QR}_i)\mathbf{e}_y)$ with $m$ denoting the local magnetization. Under this restriction Fourier transform and diagonalization lead to the variational energy per site

$$E_{SDW} = \frac{1}{L} \sum_{\mathbf{k}} (E_-(\mathbf{k}) \, n_-(\mathbf{k}) + E_+(\mathbf{k}) \, n_+(\mathbf{k})) + \frac{\Delta^2}{U} + \frac{U}{4} n^2, \qquad (3)$$

where we have used the abbreviations

$$E_\pm(\mathbf{k}) = \bar{\varepsilon}(\mathbf{k}) \pm \sqrt{\Delta^2 + \Delta\varepsilon^2(\mathbf{k})}$$

$$\left.\begin{array}{c} \bar{\varepsilon}(\mathbf{k}) \\ \Delta\varepsilon(\mathbf{k}) \end{array}\right\} = \frac{\varepsilon(\mathbf{k}+\mathbf{Q}/2) \pm \varepsilon(\mathbf{k}-\mathbf{Q}/2)}{2}$$

and where we introduced $\Delta$ for $Um/2$.

Minimizing the energy with respect to $\Delta$ results in

$$\frac{1}{U} = \frac{1}{2L} \sum_{\mathbf{k}} \frac{n_-(\mathbf{k}) - n_+(\mathbf{k})}{\sqrt{\Delta^2 + \Delta\varepsilon^2(\mathbf{k})}} \qquad (4)$$

which is equivalent to the fulfilment of the self-consistency condition $m = 2|\langle \mathbf{S} \rangle|$. In the limit $\Delta \to 0$ the right hand side of this equation represents the paramagnetic susceptibility $\chi(\mathbf{Q})$ and the critical $U$ at which the paramagnetic state becomes unstable against a second order transition to a helical configuration with twist vector $\mathbf{Q}$ is given by

$$\frac{1}{U_{cr}} = \chi(\mathbf{Q}) = \frac{1}{L} \sum_{\mathbf{k}} \frac{n(\mathbf{k}-\mathbf{Q}/2) - n(\mathbf{k}+\mathbf{Q}/2)}{\varepsilon(\mathbf{k}+\mathbf{Q}/2) - \varepsilon(\mathbf{k}-\mathbf{Q}/2)}. \qquad (5)$$

We now turn to the special case of the triangular lattice for which we obtain the phase diagram presented in Fig. 2. The boundary between the paramagnetic phase and helical configurations is indicated by the lower full line. We find that the instability of the paramagnetic state occurs for those $\mathbf{Q}$-vectors for which the two Fermi surfaces $\varepsilon(\mathbf{k} \pm \mathbf{Q}/2) = \varepsilon_F$ are just tangent to each other (imperfect nesting). With increasing particle density, the favoured $\mathbf{Q}$-vectors follow straight lines in the BZ along the para-helical phase boundary as depicted in Fig. 3: starting at the center $\mathbf{Q} = \mathbf{0}$ of the BZ for $\varepsilon_F = -6\,t$ (point A in Figs. 2 and 3) the $\mathbf{Q}$ vector reaches $\mathbf{Q} = (0, 2\pi/\sqrt{3})$ on the boundary of the BZ for $\varepsilon_F = -2\,t$ (point B), moves along the boundary and reaches the corner $\mathbf{Q} = (2\pi/3, 2\pi/\sqrt{3})$ of the BZ for $\varepsilon_F = -t$ (point C) and then turns back towards the center as $\varepsilon_F = 2\,t$ is approached. In the limit of $\varepsilon_F \to 2\,t$ (which corresponds to $n \to 1.5$) the Stoner criterion tells us that the critical coupling for $\mathbf{Q} = \mathbf{0}$ approaches zero due to the logarithmic van Hove singularity: $U_{cr}^{-1} = \rho_\triangle(\varepsilon_F) \propto -3/(4\pi^2 t) \ln|2 - \varepsilon_F/t|$. However, this ferromagnetic instability is not the leading one, because the Fermi surface is a perfectly nested hexagon. The corresponding nesting vectors $\mathbf{Q} = (0, 2\pi/\sqrt{3}), (\pm\pi, \pi/\sqrt{3})$ all correspond to alternating rows of up and down spins. This structure will be called the Néel structure in what follows.

Due to this nesting the favoured $\mathbf{Q}$ vector right before $\varepsilon_F = 2\ t$ jumps from a value near zero to one of the nesting vectors for which $U_{cr}$ then vanishes as $tU_{cr}^{-1} \propto (\ln|2 - \varepsilon_F/t|)^2$, i.e. it drops faster than corresponding to the Stoner criterion. For $\varepsilon_F > 2\ t$ the favoured $\mathbf{Q}$ first moves from one of the nesting vectors towards $\mathbf{Q} = \mathbf{0}$, jumps onto the line A-C for even higher $\varepsilon_F$ and finally returns towards $\mathbf{Q} = \mathbf{0}$.

For Coulomb interactions above the instability of the paramagnetic state we find areas in the $(U, n)$ phase diagram where the optimum $\mathbf{Q}$ varies continuously as well as other areas where the optimum $\mathbf{Q}$ has a fixed value such that particular spin patterns are stable in a whole area. The stable patterns are
i) the ferromagnetic spin pattern $\mathbf{Q} = \mathbf{0}$,
ii) the antiferromagnetic three sublattice 120° structure $\mathbf{Q}_{AF} = (4\pi/3, 0)$,
iii) the Néel structure of alternating rows of up and down spins $\mathbf{Q}_{\text{Néel}} = (0, 2\pi/\sqrt{3})$.

Let us now briefly discuss the different stability regions starting with the antiferromagnetic 120° structure. At half filling this structure is favoured for all $U$ exceeding $5.27\ t$ (see [20, 21] for a more detailed discussion). In the limit of large $U$ the stability of this structure follows easily from the asymptotic expression for the SDW energy

$$E_{SDW}(n=1) = -\frac{zt^2 + \varepsilon(\mathbf{Q})t}{2U} \geq -\frac{zt^2 + \varepsilon(\mathbf{Q}_{AF})t}{2U} = -\frac{Wt}{2U},$$

$$m = 1 - \frac{zt^2 + \varepsilon(\mathbf{Q})t}{U^2}$$

with $W$ denoting the band width ($W_\triangle = 9t$).
The 120° alignment remains favoured if the density is lowered moderately ($n < 1$). To see this we start at half filling and remove one single particle ($U$ shall be large enough to avoid band overlap). The energy changes by

$$
\begin{aligned}
dE &= \min_{\mathbf{k}}(-E_-(\mathbf{k}) - \frac{U}{2}n) = \min_{\mathbf{k}}(-\bar{\varepsilon}(\mathbf{k}) + \sqrt{\Delta^2 + \Delta\varepsilon^2(\mathbf{k})} - \frac{U}{2}) \\
&\geq -3t - \frac{U}{2}(1-m).
\end{aligned}
\tag{6}
$$

Equality only holds for the ferromagnetic structure and for the 120°-structure. As the latter structure has the lowest magnetization for given $U$ it is not only favoured at $n=1$ but also gains most energy if one particle is removed. Therefore the AF is stable in a finite range of densities $n \leq 1$ and the favoured twist vector doesn't change with $n$. Decreasing $n$ further we find a first order transition to the Nagaoka state (for large $U$), whereas for lower $U$ there is a first order transition to the Néel-structure.

If, on the other hand, the density is increased above half filling the 120° structure immediately gets unstable and the favoured $\mathbf{Q}$ continuously turns towards $\mathbf{Q} = \mathbf{0}$. These intermediate helical states are however not really stable for large $U$ as the corresponding energy is not convex as a function of the density $n$. Therefore the system is unstable against a phase separation between the antiferromagnetic and the Nagaoka phase and one has to use the Maxwellian construction to restore convexity. The critical density $n_{cr}$ below which phase separation takes place in dependence of the Coulomb coupling follows from the equation

$$
E_{SDW}(U, n=1) = E_{\mathcal{N}}(n_{cr}) - E'_{\mathcal{N}}(n_{cr})(n_{cr} - 1),
\tag{7}
$$

where $E_{\mathcal{N}}$ denotes the energy of the Nagaoka phase. The asymptotic evaluation for $U \to \infty$ of (7) leads to $t/U = 4\pi\,(n_{cr}-1)^2/(3\sqrt{3})$. This is the strongest reduction of the stability range of the Nagaoka state in the limit $U \to \infty$, we were able to achieve and will not be exceeded in the following sections. Especially, it is clearly superior to the linear asymptotic behaviour $t/U \propto n_{cr} - 1$ which would be the result of analysing the instability of the Nagaoka state against a vanishing twist $\partial^2 E_{SDW}/\partial Q_i^2|_{\mathbf{Q}=\mathbf{0}} = 0$. This last instability gets relevant, however, if one lowers $U$ as the corresponding curve crosses the phase separation curve discussed above. Below the crossing point phase separation takes place between the 120° structure and helical states $\mathbf{Q} = (q_1, 0)$ and the instability line of the Nagaoka state is determined by $\partial^2 E_{SDW}/\partial Q_i^2|_{\mathbf{Q}=\mathbf{0}} = 0$.

Of course, the Nagaoka state might also be unstable against a partially polarized ferromagnetic state. If depolarization is supposed to be a continuous process it can be studied in terms of a single spin flip problem. The obvious criterion is that the gain of band energy due to the flipped spin $\varepsilon_t - \varepsilon_F$ should equal the cost in Coulomb energy $U(2-n)$. The explicit calculation shows that this HF single spin flip instability is never the relevant criterion, i.e. we found no regions in the phase diagram where a less than fully polarized ferromagnetic state is stable.

So much for the HF phase diagram of the Hubbard model on the triangular lattice as far as helical SDW states are considered. Summarizing our results concerning the Nagaoka state one has to realize that its stability region is clearly overestimated by the HF approximation. In particular, HF theory doesn't exclude stability for any filling factor in the limit $U \to \infty$.

A well-known ansatz for improving HF theory uses a Gutzwiller projector controlling double occupancy. Unfortunately, except for high dimensions [24, 25] a rigorous evaluation of the energy of Gutzwiller wave functions is not possible. It is, though, possible for the very special but important case of a Nagaoka state with only one spin flipped. Gutzwiller single spin flip wave functions were used by SKA [10]. The results of this work will be the starting point of our next subsection.

## 2.2 Variational single spin flip states

### 2.2.1 Formulation

The variational single spin flip states which will be discussed in the following were formulated in order to achieve an energy lower than the Nagaoka energy and thereby prove the instability of the Nagaoka state in a certain density range. The general method consists of minimizing the energy of the spin flip state $|SF\rangle$ with respect to the variational parameters which are contained in the trial wave function. From the condition

$$\Delta e = \frac{\langle SF|\mathcal{H}|SF\rangle}{\langle SF|SF\rangle} - E_\mathcal{N} < 0 \quad , \tag{8}$$

with $E_\mathcal{N}$ being the Nagaoka energy, one attains an area in the phase diagram where the Nagaoka state is definitely unstable. Whenever $\Delta e > 0$ the Nagaoka state might be a ground state.

According to the structure of the Hubbard Hamiltonian $\Delta e$ contains three parts: the change in kinetic energy of the spin up particles due to the spin flip, the kinetic energy the down spin gains by delocalizing and the potential energy which results from double occupancies made possible by the spin flip. Under the restriction that the particles involved in the spin flip are given definite momenta the lowest possible energy is achieved by removing the spin up particle from the Fermi edge and creating the spin down particle with a momentum $\mathbf{q}_b$ belonging to the lower band edge $\varepsilon_b$.

For our calculations we always assume $n < 1$. This is possible because the particle hole transformation maps the case $n > 1$ to less than half filling by changing the sign of $t$. In the phase diagrams for the non-bipartite lattices, however, we show the full density range $0 < n < 2$ for positive $t$.

### 2.2.2 SKA Gutzwiller wave function

The variational ansatz originally considered by Shastry, Krishnamurthy and Anderson [10] consists of a Gutzwiller projector applied to a single spin flip state constructed as explained above

$$|SKA\rangle = \prod_{\mathbf{l}} \left(1 - (1-g)\, n_{\mathbf{l}\uparrow} n_{\mathbf{l}\downarrow}\right) c^+_{\mathbf{q}_b \downarrow} c_{\mathbf{k}_F \uparrow} |\mathcal{N}\rangle \qquad (9)$$

$$= \frac{1}{\sqrt{L}} \sum_{\mathbf{l}} e^{i\mathbf{q}_b \mathbf{l}} (1 - (1-g) n_{\mathbf{l}\uparrow}) c^+_{\mathbf{l}\downarrow} c_{\mathbf{k}_F \uparrow} |\mathcal{N}\rangle$$

with the variational Gutzwiller parameter $g$ controlling the probability of double occupancies. For $g = 1$ correlations are not taken into account and we recover the rather trivial single spin flip problem already mentioned in the HF part of our paper. The case $U = \infty$, however, corresponds to $g = 0$ when double occupancies are completely prohibited.

The spin flip energy $\Delta e_\infty$ for infinite $U$ is calculated to be [10]

$$\Delta e_\infty = e_\uparrow + e_\downarrow \quad , \quad e_\uparrow = -(e_\mathcal{N}/\delta) - \varepsilon_F \quad , \quad e_\downarrow = \varepsilon_b\, \delta\, (1 - (e_\mathcal{N}/\delta z t)^2) \quad (10)$$

with the lattice coordination number $z$, the Nagaoka energy per lattice site $e_\mathcal{N}$ and the hole density $\delta = 1 - n$. There is no potential energy due to the absence of double occupancies. Evaluating $\Delta e = 0$ for finite $U$ by optimizing $g$ for each given $U$ and $\delta$ leads to an analytic expression for the Nagaoka instability line $U_{cr}(\delta)$ [14]:

$$U_{cr}(\delta) = \frac{(\varepsilon_F - \varepsilon_b)[\Delta e_\infty + (\varepsilon_F - \varepsilon_b)\delta]}{(1-\delta)\Delta e_\infty} \quad . \qquad (11)$$

The lattice structure enters into (11) via the density of states which is needed to calculate $\delta$ and $e_\mathcal{N}$ as functions of the Fermi energy.

In Fig. 4 we show the instability line obtained for the triangular lattice. As expected the Nagaoka state turns out to be completely unstable for less than half filling, whereas for more than half filling the spin flip energy remains always positive for large $U$. In order to demonstrate the importance of the Gutzwiller projector we also show again the HF result from Fig. 2 obtained by setting $g = 1$ which leads to a gross overestimate of the stability range of the Nagaoka state, especially for less than half filling.

### 2.2.3 Including majority spin hopping processes

An extended single spin flip ansatz was introduced by Basile and Elser in 1990 [11] to investigate the instability of the Nagaoka state at infinite $U$. In order to lower the kinetic energy loss of the spin up particles due to the existence of the flipped spin they extended the SKA Gutzwiller wave function including spin up hopping terms. These terms allow the spin up particles to move away from the down spin position $\mathbf{l}$ to other lattice sites $\mathbf{r}$ instead of simply projecting out double occupancies:

$$|BE^\infty\rangle = \frac{1}{\sqrt{L}} \sum_{\mathbf{l}} e^{i\mathbf{q}_b \mathbf{l}}\, c_{\mathbf{l}\uparrow} \sum_{\mathbf{r}} f_{\mathbf{r}-\mathbf{l}}\, c^+_{\mathbf{r}\uparrow}\, c^+_{\mathbf{l}\downarrow}\, c_{\mathbf{k}_F\uparrow}|\mathcal{N}\rangle \quad . \tag{12}$$

The amplitudes of the hopping processes are controlled by the variational parameters $f_{\mathbf{r}-\mathbf{l}}$ which are assumed to reflect the symmetry of the underlying lattice.

Basile and Elser investigated this trial state numerically for a finite square lattice with 61 sites. Hanisch and Müller-Hartmann [14] used a modification of this ansatz making it suitable for an evaluation in the thermodynamic limit and for all values of $U$. Starting from nearest neighbour hopping they successively added further spin up hopping processes. In order to consider finite $U$ double occupancies were made possible by including the term $g \cdot n_{\mathbf{l}\uparrow}$ with the Gutzwiller parameter already known from the SKA wave function (9). The Basile-Elser like ansatz for nearest neighbour hopping only is given by

$$|BE_{nn}\rangle = \frac{1}{\sqrt{L}} \sum_{\mathbf{l}} e^{i\mathbf{q}_b \mathbf{l}} \left[ (1 - (1-g)n_{\mathbf{l}\uparrow}) + f_1 c_{\mathbf{l}\uparrow} \sum_{\mathbf{n}} c^+_{\mathbf{n}\uparrow} \right] c^+_{\mathbf{l}\downarrow} c_{\mathbf{k}_F\uparrow}|\mathcal{N}\rangle \quad . \tag{13}$$

The amplitude of the hopping processes to the neighbouring sites $\mathbf{n}$ of the down spin is controlled by the variational parameter $f_{\mathbf{n}-\mathbf{l}} \equiv f_1$, the variational parameter $f_\mathbf{0}$ has been set to 1. The lattice structure enters into the calculation of the spin flip energy for $|BE_{nn}\rangle$ again only via the density of states. As a consequence this ansatz allows the investigation of the instability of the Nagaoka state in the thermodynamic limit on an arbitrary lattice without great numerical effort. We will apply (13) also to the honeycomb and kagome lattices in sections 3 and 4, respectively.

Going beyond this ansatz by including further hopping processes we obtain the spin flip energy in general as a rational function in the elements of the single particle density matrix. In [14] Hanisch and Müller–Hartmann evaluated these matrix elements for the square lattice. For the present calculations we have developed a method to calculate the elements of the single particle density matrix for the triangular lattice efficiently such that only one numerical integration has to be performed for each matrix element. As described in the Appendix we used recursion formulae to express the integrand in terms of complete elliptic integrals of the first and second kind. The minimization of the spin flip energy with respect to the variational parameters corresponds to the solution of a generalized eigenvalue problem.

The long-dashed line in Fig. 4 shows the Nagaoka instability according to the Basile–Elser criterion for the triangular lattice including nearest neighbour hopping only. For more than half filling the SKA result is significantly improved, in particular in the region of large $U$ and $n$. For the first time it becomes obvious that the Nagaoka state is not stable for all $n > 1$ at infinite $U$. The critical density above which the Nagaoka state cannot represent the ground state of the Hubbard model is calculated to be $n_{cr} = 1.912$. An even lower value of $n_{cr} = 1.887$ is achieved if further hopping processes are taken into account. This decrease of the critical density is achieved by including into the variational ansatz spin up hopping to the 6 first, 6 second and 12 fourth neighbours. Including hopping to even more distant neighbours does not affect the result within the numerical accuracy as the importance of the processes diminishes with increasing distance from the flipped spin. The 6 third neighbours are an exception in this context since they do not influence the result for $n_{cr}$.

As already mentioned the improvements achieved by the Basile–Elser ansatz in comparison to SKA are most pronounced near $n = 2$. In the $t = -1$ picture for less than half filling, which we used for our calculations, this corresponds to the range of large hole densities where the hopping processes are most efficient. For small hole densities most of the sites near the down spin position are occupied and therefore few hopping processes are allowed, especially for large $U$ where this extension of the SKA ansatz therefore has only a small effect (see Fig. 4). In this region of the phase diagram, however, a distinct improvement of the SKA result is achieved by allowing the system to gain the antiferromagnetic exchange energy of the order $t^2/U$ between the down spin and the neighbouring up spins. This is done by replacing the Gutzwiller projector in (13) with another term allowing double occupancies, namely introducing a simple local correlation effect by binding a hole to the neighbours of the doubly occupied site:

$$1 - (1-g)n_{\mathbf{l}\uparrow} \longrightarrow (1 - n_{\mathbf{l}\uparrow}) + \alpha \sum_{\mathbf{m}} (1 - n_{\mathbf{m}\uparrow}) \,. \qquad (14)$$

Here $\underline{m}$ stands for the nearest neighbour sites of the down spin site $\mathbf{l}$ and $\alpha$ is a variational parameter. The full line of Fig. 4 marks the Nagaoka instability line obtained with a variational single spin flip state including the Basile-Elser hopping processes mentioned before and the exchange term (14). As for the square lattice [14] this ansatz leads to a qualitatively better result than the Basile-Elser wave function with a simple Gutzwiller projector and correctly reproduces the divergence of the critical coupling $U_{cr}$ in the limit $n \searrow 1$. The critical density $n_{cr} = 1.887$ is not influenced by the inclusion of the exchange term while the minimal critical coupling is slightly improved to $U_{cr} = 9.13\ t$ which is the value at $n = 1.586$.

We also considered another local correlation term introduced in [14] which makes the amplitude for the creation of a down spin at the lattice site **l** depend on the spin up occupancy of the nearest neighbours **r** of **l** in order to improve the spin down kinetic energy. This effect is included in the variational ansatz by extending (14)

$$(1 - n_{\mathbf{l}\uparrow}) \longrightarrow (1 - n_{\mathbf{l}\uparrow})(1 + \beta \sum_{\mathbf{r}}(1 - n_{\mathbf{r}\uparrow})) \tag{15}$$

with an additional variational parameter $\beta$. For the square lattice, this leads to a pronounced reduction of the critical hole density [14] compared with the SKA and BE results whereas for the triangular lattice it does not improve at all the value of $n_{cr}$ obtained with the Basile-Elser like ansatz. The physical reason for this is the fact that (again in the $t = -1$ picture for less than half filling) the hole densities being in consideration for the Nagaoka instability at infinite $U$ for the triangular lattice are much larger than for the square lattice and consequently making an additional restriction by projecting out double occupancies at the neighbouring sites of the down spin position is inefficient since most of these sites are unoccupied anyway. For intermediate hole densities the term is also not relevant since the critical coupling $U_{cr}$ is so small that the system prefers the moderate potential energy caused by double occupancies rather than the loss of spin up kinetic energy due to the strict suppression of double occupancies near the flipped spin contained in (15).

## 2.3 A static scattering problem

In the limit of small hole densities the Nagaoka state was shown to be unstable for all finite values of $U$ on arbitrary lattices already in the early work of Richmond and Rickayzen [26]. In their variational single spin flip state the down spin is created at a fixed lattice site and this way represents a static scattering potential for the spin up particles. The kinetic energy of the flipped spin is neglected which is a good approximation for large $U$ and small $\delta$ because under these conditions the spin down particle is unlikely to move.

The spin up particles no longer occupy Bloch states but scattering states and the calculation of the energy spectrum is reduced to a single particle problem. Richmond and Rickayzen used the Green's function method to derive an analytic formula for the energy difference between the Nagaoka state and their variational state:

$$\Delta e = \frac{1}{\pi} \int_{\varepsilon_b}^{\varepsilon_F} d\varepsilon \ \arctan\left(\frac{\pi U \rho(\varepsilon)}{1 - U\Delta(\varepsilon)}\right) - \varepsilon_F \ . \qquad (16)$$

Here $\Delta$ and $\pi\rho$ stand for the real and the imaginary part of the local propagator, respectively. $\Delta$ is calculated from the density of states $\rho$ by using a Kramers–Kronig relation.

Richmond and Rickayzen achieved their general results by evaluating (16) in the asymptotic limit of $U \to \infty$. In our present work we determined the Nagaoka instability line for the triangular lattice according to the Richmond–Rickayzen criterion exactly by numerical evaluation of $\Delta e = 0$ for all $1 < n < 2$. We want to remind the reader that the limit $n \searrow 1$ is mapped to the Richmond-Rickayzen limit $\delta \searrow 0$ by changing the sign of $t$.

To evaluate (16) completely for all values of $n$ a double numerical integration is necessary. In purpose to diminish the numerical inaccuracies it is useful to treat singularities in an analytical way. As the density of states $\rho_\triangle(\varepsilon)$ is known to diverge logarithmically at $\varepsilon = 2t$ it is possible to find a regularized expression for $\rho_\triangle$. By inserting this regularized form of the integrand into the Kramers–Kronig relation one part of the integral can be expressed explicitly in terms of dilogarithm functions. This way we showed that the real part $\Delta$ of the local propagator has a discontinuity by $\frac{3}{4|t|}$ at $\varepsilon = 2t$. The integrands of the remaining integrals which have to be solved numerically are all continuous functions.

In Fig. 5 our result for the Richmond-Rickayzen state is compared with the results of other variational criteria which have turned out to be reasonable near half filling (i. e. $n \searrow 1$), namely the single spin flip state (14) including exchange processes and the two curves obtained in HF theory by evaluating the instability of the Nagaoka state against an infinitesimal twist and against phase separation, respectively. The Richmond-Rickayzen criterion is the best local instability criterion in the limit $\delta \to 0$ since $U_{cr}^{-1} \propto \delta/\ln(1/\delta)$. The global instability due to phase separation, however, leads to a substantially stronger criterion since it is the only one where the critical coupling diverges as $U_{cr}^{-1} \propto \delta^2$.

For larger hole densities the Richmond-Rickayzen ansatz is a poor variational state due to the fact that there the kinetic energy of the spin down particle plays an important role in reducing the energy of the spin flip state.

Beyond $\delta(\varepsilon_F = 0) = 0.399$ the Richmond-Rickayzen ansatz is not sufficient for infering the instability of the Nagaoka state for any $U > 0$ (see Fig. 5).

## 2.4 Global instability of the Nagaoka state

As we have already seen by evaluating the SKA ansatz the Nagaoka state is locally unstable for less than half filling on the triangular lattice. In particular, for $(1 - n) \ll 1$ and $U = \infty$ the SKA wave function leads to an energy gain of

$$\begin{aligned}\Delta e &= (1-n)\,[\rho^{-1}(\varepsilon_t)/2 - z\,t\,(1 - \varepsilon_t^2/(zt)^2)] \\ &= (1-n)\,t\,(\pi\sqrt{3} - 9)/2 \approx -1.78\,t\,(1-n).\end{aligned}$$

In this section we will present a general consideration which shows that a single hole produces a global instability of the Nagaoka state.

Let $U = \infty$, $t > 0$ and consider a lattice of $L$ sites with periodic boundary conditions and $N = L - 1$ electrons (one hole). We are going to show that whenever there are at least two inequivalent points $\mathbf{k}$ in the BZ with $\varepsilon(\mathbf{k}) = \varepsilon_t$ then there exist helical states with energy $E < E_\mathcal{N} = -\varepsilon_t$.

The finite energy gain $dE = E - E_\mathcal{N}$ (independent of $L$) makes sure that the instability also holds for (possibly very small but) finite hole densities in the thermodynamic limit. Therefore our proof is of thermodynamic relevance. As a by-product the explicit construction of such states provides information about which magnetic structures are favoured by the hole hopping although we are not able to treat the paramagnet in an adequate manner.

To prove the above statement we start from an arbitrary (HF) helical spin structure:

$$|\mathbf{Q}\rangle = \prod_i \left(\cos(\frac{\mathbf{QR}_i}{2})c_{i\uparrow}^+ - \sin(\frac{\mathbf{QR}_i}{2})c_{i\downarrow}^+\right)|0\rangle.$$

To proceed it is useful to define holon and doublon operators in such a way that the above helical structure corresponds to a fermion vacuum:

$$h_i = \cos(\frac{\mathbf{QR}_i}{2})c_{i\uparrow}^+ - \sin(\frac{\mathbf{QR}_i}{2})c_{i\downarrow}^+, \quad d_i = \sin(\frac{\mathbf{QR}_i}{2})c_{i\uparrow} + \cos(\frac{\mathbf{QR}_i}{2})c_{i\downarrow}.$$

The hopping part of the Hubbard Hamiltonian then reads ($\mathbf{R}_{ij} := \mathbf{R}_j - \mathbf{R}_i$)

$$\mathcal{H}_t = -t\sum_{\langle i,j\rangle}[\cos(\frac{\mathbf{QR}_{ij}}{2})(h_i h_j^+ + d_i^+ d_j) + \sin(\frac{\mathbf{QR}_{ij}}{2})(h_i d_j - d_i^+ h_j^+) + h.c.].$$

Now one easily calculates the energy of the following superposition of one hole states
$$|\mathbf{k}\rangle_{\mathbf{Q}} = h_{\mathbf{k}}^{+}|\mathbf{Q}\rangle = L^{-1/2}\sum_{i}\exp(i\mathbf{k}\mathbf{R}_{i})h_{i}^{+}|\mathbf{Q}\rangle$$
to be
$$\varepsilon_{L}^{(1)}(\mathbf{k},\mathbf{Q}) = -(\varepsilon(\mathbf{k}+\mathbf{Q}/2)+\varepsilon(\mathbf{k}-\mathbf{Q}/2))/2 = -\bar{\varepsilon}(\mathbf{k})$$
which reproduces equation (6) in the limit $U \to \infty$. Defining $\varepsilon_{L}^{(1)}(\mathbf{Q}) = \min_{\mathbf{k}}\varepsilon_{L}^{(1)}(\mathbf{k},\mathbf{Q})$ we obtain the energy $\varepsilon_{L}^{(1)}(\mathbf{0}) = -\varepsilon_{t}$ for the Nagaoka state. If there are two band maxima at inequivalent points $\mathbf{k}_1$ and $\mathbf{k}_2$ in the BZ, we obtain the same energy $\varepsilon_{L}^{(1)}(\mathbf{Q}) = -\varepsilon_{t}$ for a helical state with $\mathbf{Q} = \mathbf{k}_1 - \mathbf{k}_2 \neq \mathbf{0}$. However, this state is not an eigenstate of the hole hopping Hamiltonian $\mathcal{P}\mathcal{H}_t\mathcal{P}$ ($\mathcal{P}$ denotes the projector onto states without double occupancies). Applying the analytic Lanczos method to this state we can therefore construct helical one hole states with energies below $-\varepsilon_t$. Performing the first Lanczos step explicitly we obtain the state

$$|k\rangle_{\mathbf{Q}} = L^{-1/2}\sum_{i}\exp(i\mathbf{k}\mathbf{R}_i)h_i^{+}(\alpha + \sum_{j}\beta_{j-i}d_j^{+}h_j^{+})|\mathbf{Q}\rangle$$

(flipped spin at site $j$ neighboured to the hole at site i). Assuming the normalization $|\alpha|^2 + \sum_j |\beta_{j-i}|^2 = 1$ we obtain the trial energy

$$\varepsilon_L^{(2)}(\mathbf{k},\mathbf{Q}) = -|\alpha|^2\bar{\varepsilon}(\mathbf{k}) - \sum_j t\sin(\frac{\mathbf{Q}\mathbf{R}_{ij}}{2})(\exp(-i\mathbf{k}\mathbf{R}_{ij})\alpha\beta_{i-j}^{*}+c.c.)$$
$$+\sum_j \exp(-i\mathbf{k}\mathbf{R}_{ij})\,t\cos(\frac{\mathbf{Q}\mathbf{R}_{ij}}{2})(\beta_{j-i}\beta_{i-j}^{*}+\sum_p \beta_{p-i}\beta_{p-j}^{*})$$

The sum over $p$ covers all common nearest neighbours of the neighboured pair of sites $i$ and $j$. For the triangular lattice the number of such sites $p$ for given $i$ and $j$ amounts to $z_1 = 2$.

The evaluation of $\varepsilon_L^{(2)}(\mathbf{k},\mathbf{Q})$ is especially transparent for the choice $\beta_{j-i} = \beta\exp(i(\mathbf{k}-\mathbf{Q}/2)\mathbf{R}_{ij})/\sqrt{z}$, where one obtains

$$\varepsilon_L^{(2)}(\mathbf{k},\mathbf{Q}) = (\alpha^{*},\beta^{*})\begin{pmatrix} -\bar{\varepsilon}(\mathbf{k}) & i\frac{\Delta\varepsilon(\mathbf{Q}/2)}{\sqrt{z}} \\ -i\frac{\Delta\varepsilon(\mathbf{Q}/2)}{\sqrt{z}} & -\frac{\bar{\varepsilon}(\mathbf{Q}-\mathbf{k})+z_1\bar{\varepsilon}(\mathbf{Q}/2)}{z} \end{pmatrix}\begin{pmatrix} \alpha \\ \beta \end{pmatrix}.$$

In the case of the triangular lattice this results in a hole energy of $\varepsilon_L^{(2)}(\mathbf{k}^{\star},\mathbf{Q}_{AF}) = -5/4\,\varepsilon_t$ with $\mathbf{k}^{\star} = (0, 2\pi/\sqrt{3})$. This is of course only an upper limit for the true ground state energy of one single hole; the Trugman energy $-zt$ [27] provides the best lower limit known.

# 3 Honeycomb lattice

Besides the square lattice the honeycomb lattice (see Fig. 9) is one of the few examples of bipartite lattice structures in two dimensions. The honeycomb lattice is not a Bravais lattice, however, but a triangular lattice with a basis of two lattice points. This results in two energy bands with a dispersion closely related to the dispersion (2) of the triangular lattice:

$$\varepsilon_H^2(\mathbf{k}) = t \left(3t - \varepsilon_\triangle(\mathbf{k})\right). \tag{17}$$

The density of states of the honeycomb lattice is therefore given by

$$\rho_H(\varepsilon) = \left|\frac{\varepsilon}{t}\right| \rho_\triangle\left(3t - \frac{\varepsilon^2}{t}\right) \tag{18}$$

where $\rho_\triangle(\varepsilon)$ stands for the density of states of the triangular lattice discussed in section 2. As is shown in Fig. 6 the two bands are separated by a point of vanishing density of states at $\varepsilon = 0$. The logarithmic van Hove singularities correspond to fillings of $n = 3/8$ and $n = 5/8$, respectively.

In the following we are going to investigate the stability of the Nagaoka state with respect to a single spin flip on the honeycomb lattice applying some of the variational criteria developed in section 2.2. We make use of the particle-hole symmetry and consider only the case of less than half filling. Of course the tendency towards fully polarized ferromagnetism is expected to be much weaker than on the non-bipartite triangular lattice (for more than half filling). Compared to the square lattice, which is bipartite too, there are nevertheless two important differences. Since the coordination number $z_H = 3$ is smaller than $z_\square = 4$ there exist less loops in the case of the honeycomb lattice and the length of the loops is six (see Fig. 9) in contrast to a loop length of four on the square lattice. This is important because the loops on a lattice are essential for the Nagaoka mechanism as we have already stressed in section 1. The other interesting feature is the quasi-gap in the density of states at quarter filling (for the Nagaoka state) whereas $\rho_\square(\varepsilon)$ shows a logarithmic singularity at this point.

We have calculated the spin flip energy for the SKA Gutzwiller wave function and for its extension by Basile-Elser hopping to nearest neighbour sites of the down spin position. As already mentioned in section 2.2 it is possible to express all quantities which appear in the spin flip energy (like the hole density or the Nagaoka energy) in terms of integrals over the density of states and therefore the calculations are completely analogous to those for the triangular lattice if one replaces $\rho_\triangle(\varepsilon)$ by $\rho_H(\varepsilon)$. Fig. 7 shows the regions of stability of the Nagaoka state in the phase diagram obtained with the different variational wave functions. As for the triangular lattice the simple Hartree single spin flip ansatz widely overestimates the stability of the ferromagnetic state. Including the Gutzwiller projector (SKA) we achieve a critical hole density of $\delta_{cr} = 0.802$ at $U = \infty$. The Basile-Elser ansatz for nearest neighbours splits the ferromagnetic region into two parts. We find an area of possible Nagaoka stability near half filling for large values of $U$ (similar to the Basile-Elser instability curve on the square lattice [14]) and a stability island around quarter filling, again for large $U$. Outside of these two regions the Nagaoka state is definitely unstable. The critical hole densities at infinite $U$ are given by $\delta_{cr1} = 0.383$ and $\delta_{cr2} = 0.478$ for the instability gap; the upper critical hole density is $\delta_{cr3} = 0.662$. The stability island extends to a minimum critical coupling of $U_{cr} = 16.0\,t$ at $\delta = 0.535$. As we have already observed in Fig. 5 for the triangular lattice the strongest Nagaoka instability criterion for small hole densities is again the global instability against phase separation between the Nagaoka phase and an antiferromagnetic phase (see Fig. 7).

We have applied the Basile-Elser variational criterion to many lattices in different dimensions [14, 28, 29] but only for the honeycomb lattice we obtained two separate regions in the phase diagram ($0 < \delta < 1$) where the Nagaoka state is possibly stable against a single spin flip. What is the physical background of this interesting result? First we consider the limit of small hole densities. The extension of the Nagaoka theorem by Tasaki [5] makes transparent that to achieve a sufficient condition for the existence of a fully polarized ferromagnetic ground state in the Nagaoka case (that means a half filled system plus one particle at infinite $U$) each lattice site has to be contained in a loop of length three or four. The triangular lattice and the square lattice satisfy this criterion whereas the honeycomb lattice does not. The possible lack of a Nagaoka theorem suggests that the tendency towards ferromagnetism at small hole densities should be even weaker than for the square lattice. The exact diagonalization of finite honeycomb chains with up to 30 sites by Hirsch [29] shows indeed a degeneracy of the ground state energies for different values of the total spin in the small-$\delta$ region of positive spin flip energy. As a consequence the Nagaoka state is probably not the unique ground state for small hole densities although our variational estimates of the spin flip energy are not sufficient to establish the instability.

In the context of our variational calculations the stability of the Nagaoka state around quarter filling is an exclusive effect of the quasi-gap in the density of states. When $\delta$ approaches 0.5 the Fermi energy drops very quickly with increasing hole density thereby enhancing the loss of spin up kinetic energy due to the single spin flip which always contains a term $-\varepsilon_F$ because a spin up particle is removed from the Fermi edge (see section 2.2). This of course raises the stability of the Nagaoka state. Figure 8 shows the spin flip energy for the SKA Gutzwiller ansatz and for the Basile-Elser ansatz with nearest neighbour hopping at $U = \infty$, where the effect explained above is clearly recognizable. The results of exact diagonalization [29] definitely show a maximum total spin of the ground state around $\delta = 0.5$ thereby suggesting that the stability island for the Nagaoka state resulting from our calculations is real. While we can exclude Nagaoka ferromagnetism in wide parts of the phase diagram there is a high probability for a fully polarized ferromagnetic ground state for $\delta \simeq 0.5$ and large values of $U$.

# 4 Kagome lattice

The so-called kagome lattice (kagome is the japanese word for basketwork) is the line graph of the honeycomb lattice. Generally a line graph is constructed by placing a vertex in the middle of each bond of the original lattice and then connecting two of the new vertices if the corresponding bonds have a common vertex in the original lattice [7]. It follows immediately from this construction principle that starting from a lattice with $z$ bonds emanating from each vertex the coordination number of the line graph is given by $\tilde{z} = 2\,(z-1)$. It is demonstrated in Fig. 9 how the kagome lattice is formed following the rules explained above. Since the lattice structure contains triangles it is obvious that the kagome lattice is not bipartite in contrast to the honeycomb lattice. It can, however, be decomposed into three sublattices each of them having again the kagome structure.

The kagome lattice is not only of academic interest because the kagome structure is found in real systems like the anisotropic antiferromagnet $SrCr_{8-x}Ga_{4+x}O_{19}$ where the $Cr^{3+}$ ions (with spin 3/2) are located in kagome planes [16]. Moreover the line graphs of other two-dimensional lattices like the square lattice or the triangular lattice show very complicated structures due to their high coordination numbers while the kagome lattice has only $\tilde{z} = 4$. Finally it was the kagome lattice for which Mielke presented the first thermodynamically relevant proof of the stability of the Nagaoka state in a Hubbard model [7]. The proof is based upon a common feature of all line graphs of planar lattices which is the existence of a flat energy band located at the upper band edge for positive $t$ and at the lower band edge for negative $t$, respectively. This situation already makes transparent that especially in the case of negative $t$ a very strong tendency towards ferromagnetism is to be expected because it is possible to fill particles with the same spin direction into the flat band without any additional cost in kinetic energy compared to a paramagnetic state.

For $t = -1$ the general relation between the densities of states of the original lattice (LA) and the line graph (LG) is given by [30]

$$\rho_{LG}(\varepsilon) = \alpha \; \delta(\varepsilon + 2) + (1 - \alpha) \; \rho_{LA}(\varepsilon - z + 2) \qquad (19)$$

where $\alpha$ stands for the weight of the flat band appearing at $\varepsilon = -2$. A gap between the flat band and the lower edge of the continous spectrum only exists if the original lattice is non-bipartite. The upper band edge is given by the coordination number $\tilde{z}$ of the line graph. Changing the sign of $t$ means replacing $\varepsilon$ by $-\varepsilon$ in (19) such that the flat band (and possibly the gap) then appear at the upper band edge.

If $|V| = N$ is the number of vertices of the original lattice and $|E|$ the number of edges which is equal to $zN/2$ the weight of the flat band is given by [7]

$$\alpha = \frac{|E| - |V| + \sigma}{|E|} \quad \stackrel{|V| \to \infty}{\approx} \quad 1 - \frac{2}{z} \qquad (20)$$

with $\sigma = 1$ if the original lattice is bipartite and 0 otherwise. Making use of this result and of equation (18) we obtain the density of states of the kagome lattice ($t = -1$) as

$$\rho_K(\varepsilon) = \frac{1}{3} \; \delta(\varepsilon + 2) + \frac{2}{3} \; |\varepsilon - 1| \; \rho_\triangle \left( (\varepsilon - 1)^2 - 3 \right) \; . \qquad (21)$$

This means that $\rho_K$ is composed of a delta peak at $\varepsilon = -2$ with weight $1/3$ and the two bands of the honeycomb lattice shifted in energy by $+1$ and with an amplitude reduced by a factor of $2/3$. Since the honeycomb lattice is bipartite no gap appears in this case.

As already mentioned above, for $t = -1$ the fully polarized ferromagnetic state obviously is among the possible ground states of the Hubbard model for low particle densities ($0 < n < 1/3$), i. e. in the regime where the Fermi energy is in the flat band. In this regime the Nagaoka state is not the unique ground state, however, and the degeneracy of the ground state grows exponentially with the system size. Mielke [7] used statistical arguments to show that at least in part of this density regime the Coulomb energy restricts the total spin to be extensive and therefore the system is really ferromagnetic.

We have applied variational wave functions to the kagome lattice in order to investigate the instability of the Nagaoka state with respect to a single spin flip over the whole density range $0 \leq n \leq 2$. For our calculations we again assumed $0 \leq n \leq 1$, allowing both signs for the hopping matrix element $t$. First we want to consider the case of more than half filling ($t = 1$ and $1 < n < 2$) which is mapped to $t = -1$ and $0 < n < 1$ by particle-hole transformation such that we recover the situation with the flat band at the lower band edge discussed above. As long as all electrons are filled into the flat band (i. e. for low particle densities) the Nagaoka energy is just given by $e_\mathcal{N} = \varepsilon_b n$ and the local stability of the fully polarized ferromagnet for arbitrary $U$ is obvious. The spin flip energy for the SKA ansatz at $U = \infty$ in the regime of the flat band is given by the simple expression

$$\Delta e_\infty = \frac{5}{2\delta} - \frac{3\delta}{2} - 1 > 0 \qquad \left(\frac{2}{3} \leq \delta \leq 1\right) \tag{22}$$

with the hole density $\delta = 1 - n$. Since the spin flip energy is also positive for smaller hole densities the SKA Gutzwiller ansatz doesn't provide any instability of the Nagaoka state at infinite $U$. Our variational result is of course not exact for $\delta \geq 2/3$: the true spin flip energy in the regime of the flat band is zero. The phase diagram (Fig. 11) shows a strong tendency towards fully polarized ferromagnetism also for finite $U$. Only for small $U$ and small hole densities, i. e. near half filling, we find a region of guaranteed instability of the Nagaoka state. At $\delta = 1/3$ the stability of the Nagaoka state is slightly enhanced due to the quasi-gap in the density of states taken over from the honeycomb lattice (see section 3). The critical coupling $U_{cr}$ goes to zero as the hole density approaches $2/3$ which is consistent with the stability of the Nagaoka state in the regime of the flat band found by Mielke and Tasaki. The results obtained with the SKA Gutzwiller wave function are only slightly improved by the Basile-Elser ansatz for nearest neighbours in contrast to what happened for the triangular lattice where this ansatz lead to a considerable reduction of the Nagaoka stability region. The reason is that the additional spin-up hopping processes are much less important if $U$ and $\delta$ are small because the energy cost of double occupancies is moderate and the probability for finding the neighbouring sites of the spin-down position already occupied is quite high.

As explained in section 1 one would not expect to find any Nagaoka ferromagnetism in the phase diagram for less than half filling since the kagome lattice is not bipartite. With the flat band now appearing at the upper band edge $\varepsilon_t = 2$ the Nagaoka energy for small hole densities ($\delta \leq 1/3$) is given by $e_\mathcal{N} = -\varepsilon_t \delta$. As a consequence the loss of spin up kinetic energy compared to the Nagaoka state due to a single spin flip is strictly zero because the Fermi energy is equal to $\varepsilon_t$ and the Nagaoka state is therefore locally unstable in the whole regime of the flat band. The spin flip energy for the SKA Gutzwiller ansatz ($U = \infty$) is calculated to be

$$\Delta e_\infty = e_\downarrow = -3\,\delta \; < 0 \quad (0 \leq \delta \leq 1/3). \tag{23}$$

This result holds also for the Basile-Elser ansatz for nearest neighbours because the spin-up kinetic energy cannot be further improved. Fig. 10 shows the spin flip energy for both variational wave functions at infinite $U$. The quasi-gap in the density of states now appears at $\delta = 2/3$ and leads again to a pronounced enhancement of the loss of spin up kinetic energy compared to the Nagaoka state. This effect is strong enough to produce a region of positive spin flip energy around $\delta = 2/3$ even for the Basile-Elser wave function. The Nagaoka state is definitely unstable for $\delta < 0.672$ and for $\delta > 0.727$ at $U = \infty$. With decreasing $U$ the stability area shrinks and finally disappears at $U_{cr} = 129.5 \ t$. Up to now there exist only very few examples where the variational criteria of SKA and Basile-Elser leave a region of possible stability of the Nagaoka state for $n < 1$ on non-bipartite lattices. The $fcc$-lattices in dimensions $d = 4$ and $d = 5$ show extremely narrow ferromagnetic regions for $n > 0.99$ and large $U$ [29]. We expect, however, that in this limit the Nagaoka state is globally unstable due to the results we presented in section 2.4. The situation in the case of the kagome lattice is qualitatively different and quite unique because we find a stability island well away from half filling for $n < 1$. We cannot exclude, however, that this result might be influenced by the fact that we used quite simple variational wave functions where only the density of states enters in the calculation of the spin flip energy. A numerical check by exact diagonalization is probably difficult because the system size has to be quite large in order to reach the flat band situation.

We have investigated the Nagaoka instability against phase separation for the kagome lattice, too. Due to the high ground state degeneracy of the classical Heisenberg model on the kagome lattice, it is not at all obvious how to choose the magnetic structure for which the spin density calculation at half filling is performed. In order to deal with a simple structure we took the special choice of the so called $q = 0$ phase where the basis cell consisting of three spins orientated in a 120° structure is repeated over the whole lattice. The phase separation region obtained this way is included in the phase diagram of Fig. 11.

## 5 Conclusions

We have investigated the instability of the fully polarized ferromagnetic state (Nagaoka state) on the triangular, honeycomb and kagome lattices. In combination with a previous analysis of the square lattice [14] this study gave an insight into how the possible stability range of the Nagaoka state depends on the lattice structure. In particular, we verified that more than half-filled non-bipartite lattices generally are very promising candidates for the existence of a Nagaoka region. For less than half filling we were able to exclude a fully polarized ferromagnetic ground state on the triangular lattice and reduced the possible Nagaoka stability region on the kagome lattice to a small area for large $U$ around the filling $n = 1/3$. We suppose that this behaviour is related to the existence of loops of length three in these lattices and the necessity of constructive interference if a particle or hole moves once around a loop.

Our investigation of the *triangular* lattice started with a HF approximation which gave a survey over the possible magnetic phases. We concluded from the missing convexity of the energy as a function of the density for more than half filling that the system is unstable with respect to phase separation between the antiferromagnetic $120°$ structure (favoured at half filling) and the Nagaoka phase. Evaluating this instability criterion we achieved the most pronounced reduction of the possible Nagaoka stability range for a particle density slightly above half filling. Below half filling we recognized that the $120°$ structure remains a possible ground state and competes with the Nagaoka state. Therefore we suppose that in this case the instability of the Nagaoka state might be of global character. We analysed this question in more detail presenting an intuitive but strict argument that for arbitrary non-bipartite lattices with two or more band maxima at least for small hole densities $\delta = 1 - n$ there are helical states with an energy gain per hole exceeding the Nagaoka result by a finite amount.

We continued our study of the triangular lattice performing a systematic variational analysis of the instability of the Nagaoka state with respect to a single spin flip. This instability turns out to be irrelevant due to other instabilities in the framework of HF theory. The inclusion of a Gutzwiller projector (SKA), however, allows us to exclude Nagaoka ferromagnetism alltogether below half filling and leads to a significant improvement of the HF estimates for more than half filling. Further improvement is achieved by including majority spin correlation effects into the trial wave function, namely hopping processes away from the position of the flipped spin as first introduced by Basile and Elser. An efficient calculation of the spin flip energy for this variational ansatz was possible via the extensive use of complete elliptic integrals. We showed that even for infinite Coulomb repulsion the Nagaoka state is not completely stable for more than half filling, but becomes unstable above a critical density $n_{cr} = 1.887$. Allowing the spin flip wave functions to gain the antiferromagnetic exchange energy we were able to make the critical coupling $U_{cr}$ diverge in the limit $n \searrow 1$. The same type of behaviour was also achieved by evaluating the physically different approach of localizing the flipped spin at a fixed lattice site and solving a single particle scattering problem as originally suggested by Richmond and Rickayzen. For this ansatz we obtained, in fact, the best asymptotic result in the limit of $n \searrow 1$ for a local instability of the Nagaoka state. As mentioned above, however, the instability is dominated by phase separation in this limit.

For the bipartite *honeycomb* lattice we found a Nagaoka instability for infinite $U$ above a critical hole density $\delta_{cr} = 0.802$ already by evaluating the SKA Gutzwiller trial wave function. We again improved this ansatz by Basile-Elser hopping to nearest neighbour sites and obtained the interesting result that the Nagaoka stability region splits into two parts, one situated near half filling and the other around quarter filling. The upper critical hole density is reduced to $\delta_{cr} = 0.662$. We traced the probable stability of the Nagaoka state around quarter filling – confirmed by finite cluster results – to the existence of a quasi-gap in the density of states. The Nagaoka stability against a single spin flip near half filling might be irrelevant since finite cluster results suggest a global instability with respect to a paramagnetic state in this region.

For the *kagome* lattice we demonstrated that for more than half filling there is a strong tendency towards Nagaoka ferromagnetism beyond the flat band region $5/3 < n < 2$ considered by Mielke. Consistent with the existence of the flat band the critical coupling $U_{cr}$ goes to zero as $n$ approaches $5/3$. In contrast to the results obtained for other non-bipartite lattices (triangular, fcc) we found a small ferromagnetic region for very large $U$ even below half filling. In the framework of our variational single spin flip states this result is again a consequence of the quasi-gap in the density of states occuring at $n = 1/3$ strongly favouring the stability of the Nagaoka state.

In order to achieve further improvement of the variational single spin flip results there exist many possibilities to make the trial wave functions more flexible, e.g. by including majority spin correlation effects as it was already done by two of us for the square lattice [14]. Furthermore it might be helpful to give up the restriction that the removed up spin has a definite momentum. Investigations of SKA and Basile-Elser spin wave states for multiple chains and the square lattice [31] show a significant reduction of the energy due to the formation of a bound state between the hole and the flipped spin.

## ACKNOWLEDGEMENTS

This research was performed within the scientific program of the Sonderforschungsbereich 341, supported by the Deutsche Forschungsgemeinschaft. We would like to acknowledge useful discussions with R. Hirsch, in particular on his results of excact diagonalizations of finite clusters.

# Appendix
# The single particle density matrix of the triangular lattice

The elements of the single particle density matrix of the Nagaoka state can be expressed in the following way:

$$h(\mathbf{l}) = \langle c_{\mathbf{0}\uparrow} c_{\mathbf{l}\uparrow}^+ \rangle = \int_{\varepsilon_F}^{\varepsilon_t} \rho_{m,n}(\varepsilon) d\varepsilon \qquad (A.1)$$

where the integers $m$ and $n$ represent coordinates of the vector $\mathbf{l}$ with respect to primitive lattice vectors: $\mathbf{l} = \left(m + n/2, \sqrt{3}n/2\right)$. The integration extends from the Fermi energy to the upper band edge. The functions $\rho_{m,n}(\varepsilon)$ are generalized densities of states and for the triangular lattice they take the following form:

$$\rho_{m,n}(\varepsilon) = \langle e^{i\mathbf{k}\mathbf{l}} \, \delta(\varepsilon - \varepsilon_\triangle(\mathbf{k})) \rangle_{BZ} \qquad (A.2)$$

$$= \left\langle \cos\left((2m+n)\frac{k_x}{2}\right) \cos\left(n\frac{\sqrt{3}k_y}{2}\right) \delta(\varepsilon - \varepsilon_\triangle(\mathbf{k})) \right\rangle_{BZ} .$$

The momentum average has to be taken over the first Brillouin zone. For $m = n = 0$ (A.2) obviously describes the usual density of states $\rho_\triangle(\varepsilon)$ and $h(\mathbf{l} = \mathbf{0})$ is the hole density $\delta$. The point symmetry of the triangular lattice implies

$$\rho_{m,n} = \rho_{n,m} = \rho_{m+n,-n} = \rho_{-m,-n} . \qquad (A.3)$$

We therefore need $\rho_{m,n}$ only for $m, n \geq 0$.

Using Moivre's formula for $m, n \geq 0$, $\rho_{m,n}$ can be written as:

$$\rho_{m,n} = \sum_{i=0}^{\lfloor m+n/2 \rfloor} \sum_{p=0}^{i} \sum_{j=0}^{\lfloor n/2 \rfloor} \sum_{q=0}^{j} (-1)^{p+q} \binom{2m+n}{2i} \binom{i}{p} \binom{n}{2j} \binom{j}{q} \tau_{2m+n-2p, n-2q} . \qquad (A.4)$$

Here $\tau_{u,v}$ stands for

$$\tau_{u,v}(\varepsilon) = \left\langle \cos^u\left(\frac{k_x}{2}\right) \cos^v\left(\frac{\sqrt{3}k_y}{2}\right) \delta(\varepsilon - \varepsilon_\triangle(\mathbf{k})) \right\rangle_{BZ} \qquad (A.5)$$

with $u$ and $v$ being non-negative integers, $u + v$ even. We eliminate the $k_y$-integration through the $\delta$-function. Then the $\tau_{u,v}$ attain the form of elliptic integrals in $k_x$. In the following $t = -1$ will be assumed. Carrying out the substitution $x = \cos(k_x)$ one obtains

$$\tau_{u,v} = \frac{(-1)^v}{2^{\frac{u+v+2}{2}} \pi^2} \sum_{i=0}^{v} \binom{v}{i} (-1 - \varepsilon/2)^{v-i} \, J_{-\left(\frac{u-v}{2}+i\right)} \qquad (A.6)$$

with
$$J_s = \int_{x_-}^{\min(x_+,1)} \frac{(x+1)^{-s} dx}{\xi(x)} . \tag{A.7}$$

The denominator $\xi(x)$ is the positive root of

$$\xi^2(x) = (1-x^2)(3+\varepsilon - (x-1-\frac{1}{2}\varepsilon)^2) . \tag{A.8}$$

Concerning the integration limits in (A.7) the two cases of $\varepsilon > -2$ or $\varepsilon < -2$ have to be considered separately. For values of $\varepsilon$ less than $-2$ the integration limits are given by

$$x_- = 1 + \frac{1}{2}\varepsilon - \sqrt{3+\varepsilon} \quad , \quad x_+ = 1 + \frac{1}{2}\varepsilon + \sqrt{3+\varepsilon} \quad . \tag{A.9}$$

If $\varepsilon > -2$ the upper limit is 1.

It follows from the construction of the $\tau_{u,v}$ (A.6) that the index $s$ of $J_s$ can be any integer. For negative indices the binomial theorem allows to replace the $J_{-s}$ ($s \geq 0$) by other elliptic integrals $I_r$:

$$J_{-s} = \sum_{r=0}^{s} \binom{s}{r} I_r \ (s \geq 0) \tag{A.10}$$

with

$$I_r = \int_{x_-}^{\min(x_+,1)} \frac{x^r dx}{\xi(x)} , (r \geq 0) . \tag{A.11}$$

By integrating $\frac{d}{dx}(x+1)^{-s}\xi(x)$ and $\frac{d}{dx}x^s\xi(x)$, respectively, from $x_-$ to $\min(x_+, 1)$, which are both zeros of the function $\xi(x)$, one obtains the following two recursion formulae [32]:

$$\begin{aligned} J_s &= \frac{1}{(2s-1)b_1}\Big[-2sb_0 J_{s+1} + 2(1-s)b_2 J_{s-1} + (3-2s)b_3 J_{s-2} \\ &\quad + 2(2-s)b_4 J_{s-3}\Big] \ (s \geq 1) , \end{aligned} \tag{A.12}$$

$$\begin{aligned} I_s &= \frac{1}{2(1-s)a_4}\Big[(2s-3)a_3 I_{s-1} + 2(s-2)a_2 I_{s-2} + (2s-5)a_1 I_{s-3} \\ &\quad + 2(s-3)a_0 I_{s-4}\Big] \ (s \geq 3) . \end{aligned} \tag{A.13}$$

The parameters $a_i$ and $b_i$ are the coefficients obtained if one writes $\xi^2(x)$ as a polynomial in $x$ and $x+1$, respectively:

$$\xi^2(x) = \sum_{i=0}^{4} a_i x^i = \sum_{i=0}^{4} b_i (x+1)^i \quad . \tag{A.14}$$

Note that for $s = 3$ the term containing $I_{-1}$ in (A.13) vanishes. The term containing $J_{s+1}$ in (A.12) vanishes for all $s$ since $b_0 = 0$. Due to these two peculiarities the recursion formulae (A.12) and (A.13) allow to express all $I_s$ and $J_s$ by $I_0$, $I_1$ and $I_2$. These three functions are given by linear combinations of complete elliptic integrals of the first and second kind $K[m]$ and $E[m]$, respectively (see [32]):

$$I_0 = \frac{2\pi}{z_0} K\left[z_1^2/z_0^2\right] \quad , \quad I_1 = \frac{\varepsilon\pi}{3z_0} K\left[z_1^2/z_0^2\right] \quad ,$$

$$I_2 = -z_0 E\left[z_1^2/z_0^2\right] + \frac{2}{3z_0}(\varepsilon^2/4 + 2 + \chi(\varepsilon)) K\left[z_1^2/z_0^2\right]$$

$$\text{with} \quad z_0^2 = (\max(x_+,1) - x_-)(\min(x_+,1) + 1) \quad , \qquad (A.15)$$

$$z_1^2 = (\max(x_+,1) + 1)(\min(x_+,1) - x_-) \quad ,$$

$$\chi(\varepsilon) = \begin{cases} \frac{1}{2}(1 + \varepsilon/2)^2 & \text{for} \quad \varepsilon < -2 \quad , \\ 2 + \varepsilon/2 - \sqrt{3+\varepsilon} & \text{for} \quad \varepsilon > -2 \quad . \end{cases}$$

Using the above formulae we see that the generalized densities of states $\rho_{m,n}$ are all given by linear combinations of complete elliptic integrals of the first and second kind. For the calculation of the elements of the single particle density matrix only one numerical integration has to be performed.

Equations (A.4) and (A.6) show that the density of states $\rho_\triangle$ (see Fig. 1) is proportional to $I_0$ and therefore can be expressed by a complete elliptic integral of the first kind:

$$\rho_\triangle = \frac{1}{\pi z_0} K\left[z_1^2/z_0^2\right] \quad . \qquad (A.16)$$

All the other $\rho_{m,n}$ are most conveniently obtained by implementing the above formulae using an algebraic manipulation program.

# Figure captions

**Figure 1** Density of states of the triangular lattice for $t = 1$.

**Figure 2** HF phase diagram of the triangular lattice for $t > 0$. The abbreviation AF marks the stability region of the antiferromagnetic 120° structure, $N$ stands for the collinear Néel structure, $S_1$ for the phase separation region AF-Nagaoka, whereas $S_2$ corresponds to the phase separation region AF-Helical. Hatching marks the regions where we find helical states with continuously varying **Q** vectors.

**Figure 3** Brillouin zone of the triangular lattice and illustration of the path followed by the favoured **Q** vectors along the para-helical phase boundary with increasing density $0 < n < 1.5$.

**Figure 4** Phase diagram for the triangular lattice: Nagaoka instability lines obtained with the Hartree single spin flip ansatz (dotted), the SKA Gutzwiller ansatz (long dashed), the Basile-Elser ansatz with nearest neighbour hopping (short dashed) and the best Basile-Elser like wave function including the exchange term (14) (full line), $t = 1$. Hatching marks the region where the Nagaoka state might be a ground state.

**Figure 5** Phase diagram showing the instability of the Nagaoka state on the triangular lattice in the limit $n \searrow 1$: the instability lines for the Richmond-Rickayzen ansatz (full line) and for the single spin flip wave function including the exchange term (14) (dotted) are compared to the HF criterion for the instability against infinitesimal twist (long dashed) and the global instability against phase separation (short dashed).

**Figure 6** Density of states of the honeycomb lattice for $|t| = 1$.

**Figure 7** Phase diagram for the honeycomb lattice: Nagaoka instability lines obtained with the Hartree single spin flip ansatz (dotted), the SKA Gutzwiller ansatz (short dashed) and the Basile-Elser ansatz with nearest neighbour hopping (full line). The global instability against phase separation (long dashed) is also shown. Hatching marks the regions where the Nagaoka state might be a ground state. The phase diagram for more than half filling is obtained by particle-hole transformation which means replacing $\delta$ with $n$. $U_{BR} = 12.59|t|$

**Figure 8** Spin flip energy $\Delta e_\infty$ on the honeycomb lattice at infinite $U$ for the SKA Gutzwiller ansatz (dashed) and the Basile-Elser ansatz with nearest neighbour hopping (full line). The quasi-gap in the density of states (Fig. 6) corresponds to $\delta = 0.5$.

**Figure 9** Construction of the kagome lattice as the line graph of the honeycomb lattice.

**Figure 10** Spin flip energy $\Delta e_\infty$ on the kagome lattice at infinite $U$ for the SKA Gutzwiller ansatz (dashed) and the Basile-Elser ansatz with nearest neighbour hopping (full line), for less than half filling ($n < 1$). The regime $0 \leq \delta \leq 1/3$ corresponds to the flat band at the upper band edge.

**Figure 11** Phase diagram for the kagome lattice: Nagaoka instability lines obtained with the Hartree single spin flip ansatz (dotted), the SKA Gutzwiller ansatz (short dashed) and the Basile-Elser ansatz with nearest neighbour hopping (full line). The global instability against phase separation (long dashed) is also shown. The regimes $2/3 \leq n \leq 1$ and $5/3 \leq n \leq 2$ correspond to the flat bands. Hatching marks the regions where the Nagaoka state might be a ground state. $U_{BR} = 13.80|t|$

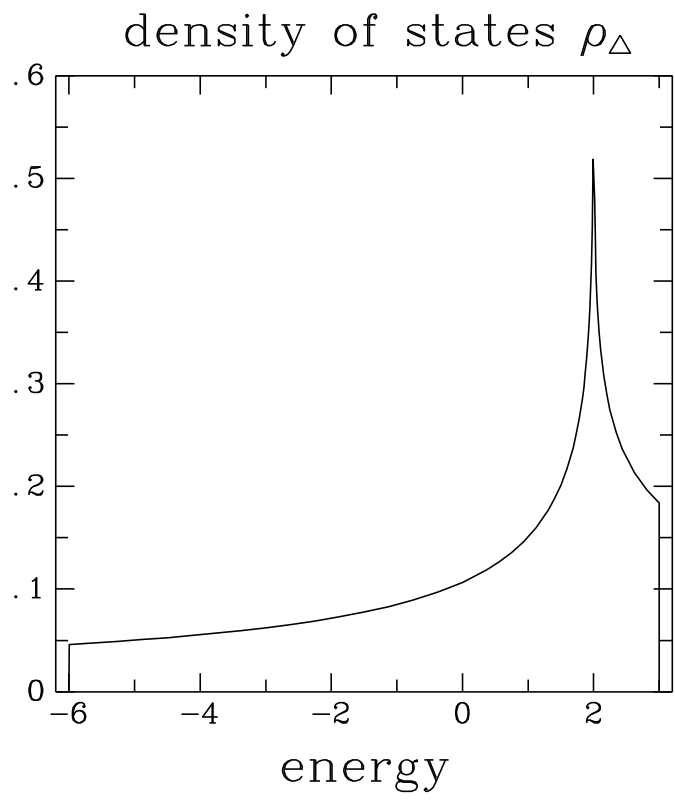

Figure 1: Density of states of the triangular lattice for $t = 1$.

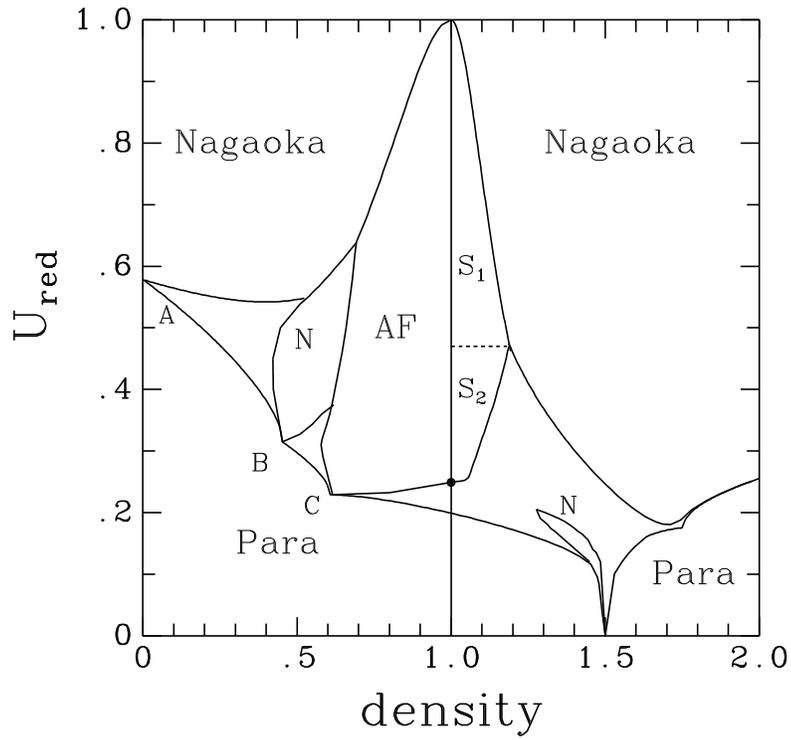

Figure 2: HF phase diagram of the triangular lattice for $t > 0$. The abbreviation AF marks the stability region of the antiferromagnetic 120° structure, $N$ stands for the collinear Néel structure, $S_1$ for the phase separation region AF-Nagaoka, whereas $S_2$ corresponds to the phase separation region AF-Helical.

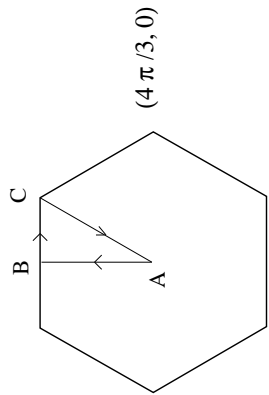

Figure 3: Brillouin zone of the triangular lattice and illustration of the path followed by the favoured **Q** vectors along the para-helical phase boundary with increasing density $0 < n < 1.5$.

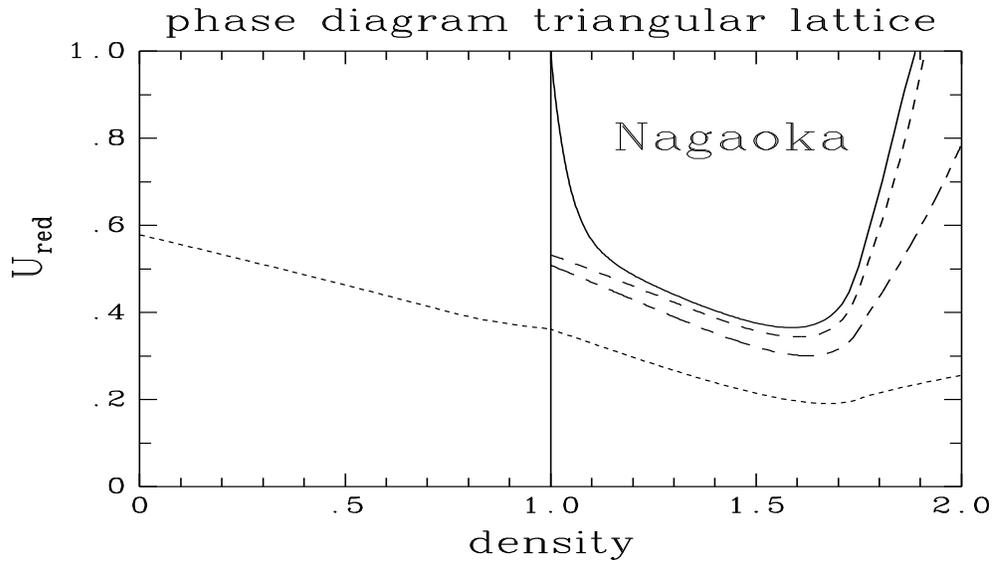

Figure 4: Phase diagram for the triangular lattice: Nagaoka instability lines obtained with the Hartree single spin flip ansatz (dotted), the SKA Gutzwiller ansatz (long dashed), the Basile-Elser ansatz with nearest neighbour hopping (short dashed) and the best Basile-Elser like wave function including the exchange term (14) (full line), $t = 1$.

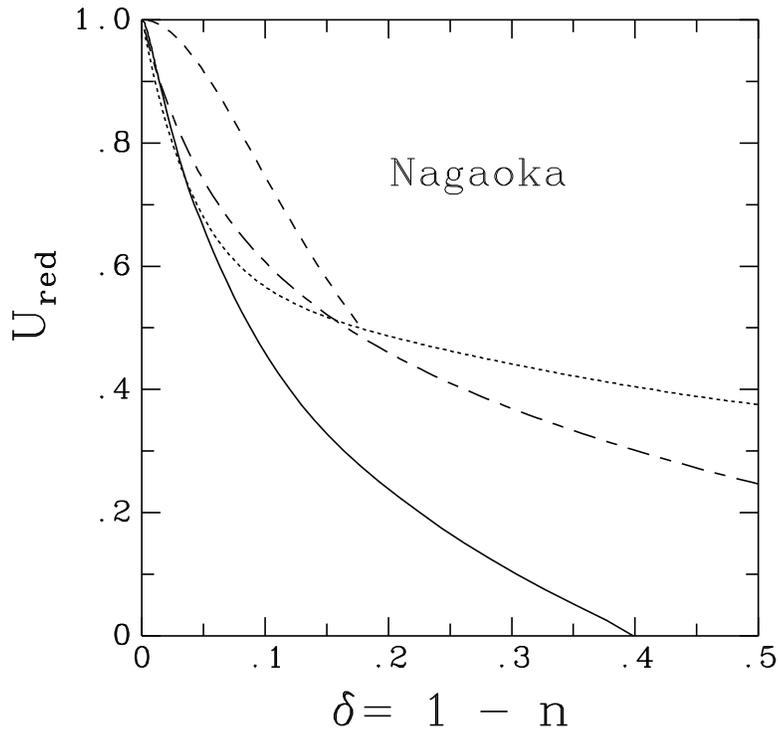

Figure 5: Phase diagram showing the instability of the Nagaoka state on the triangular lattice in the limit $n \searrow 1$: the instability lines for the Richmond-Rickayzen ansatz (full line) and for the single spin flip wave function including the exchange term (14) (dotted) are compared to the HF criterion for the instability against infinitesimal twist (long dashed) and the global instability against phase separation (short dashed).

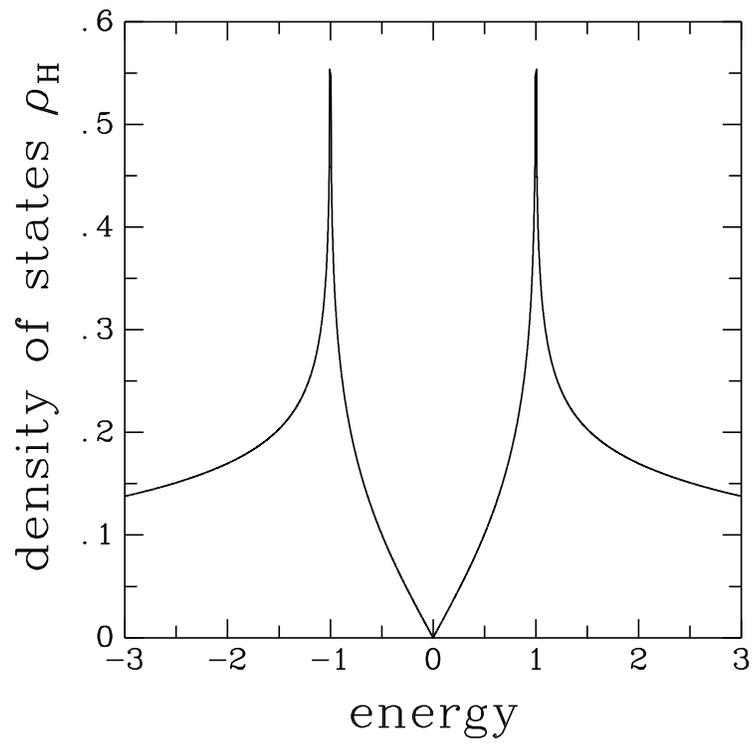

Figure 6: Density of states of the honeycomb lattice for $|t| = 1$.

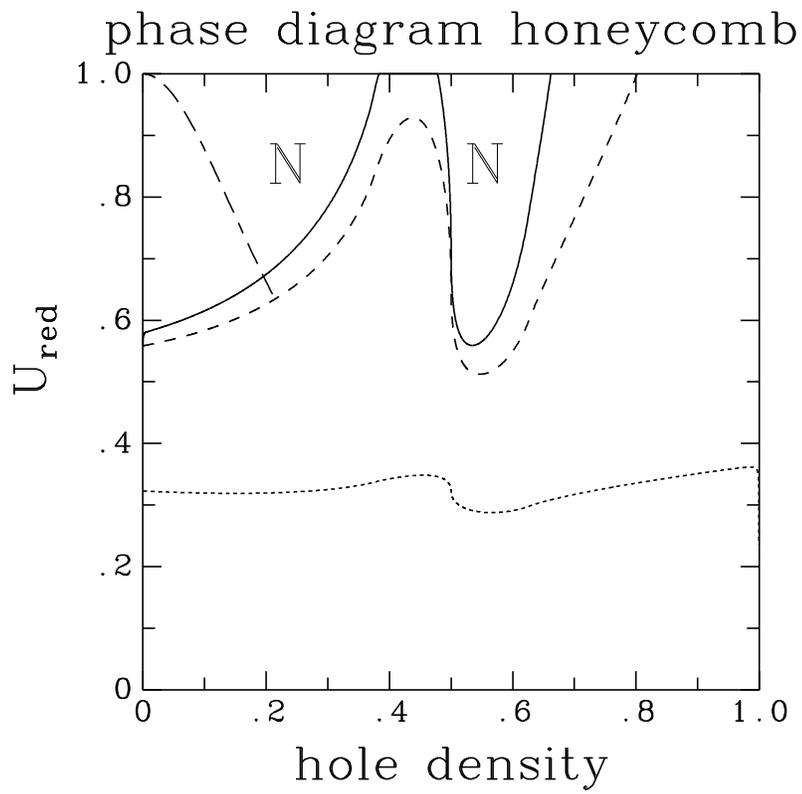

Figure 7: Phase diagram for the honeycomb lattice: Nagaoka instability lines obtained with the Hartree single spin flip ansatz (dotted), the SKA Gutzwiller ansatz (short dashed) and the Basile-Elser ansatz with nearest neighbour hopping (full line). The global instability against phase separation (long dashed) is also shown. The phase diagram for more than half filling is obtained by particle-hole transformation which means replacing $\delta$ with $n$. $U_{BR} = 12.59|t|$

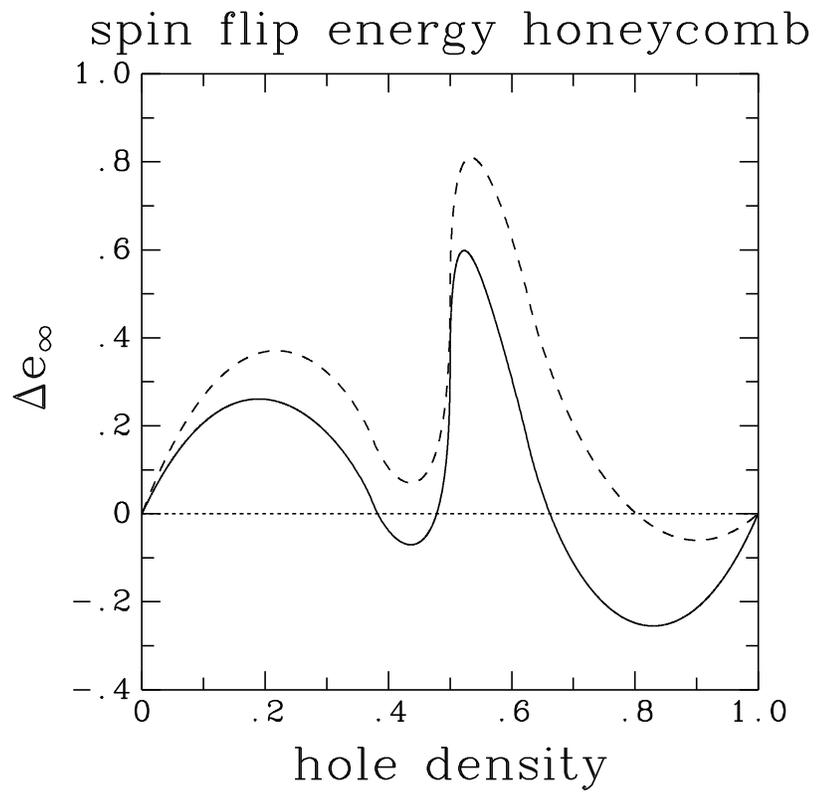

Figure 8: Spin flip energy $\Delta e_\infty$ on the honeycomb lattice at infinite $U$ for the SKA Gutzwiller ansatz (dashed) and the Basile-Elser ansatz with nearest neighbour hopping (full line). The quasi-gap in the density of states (Fig. 6) corresponds to $\delta = 0.5$.

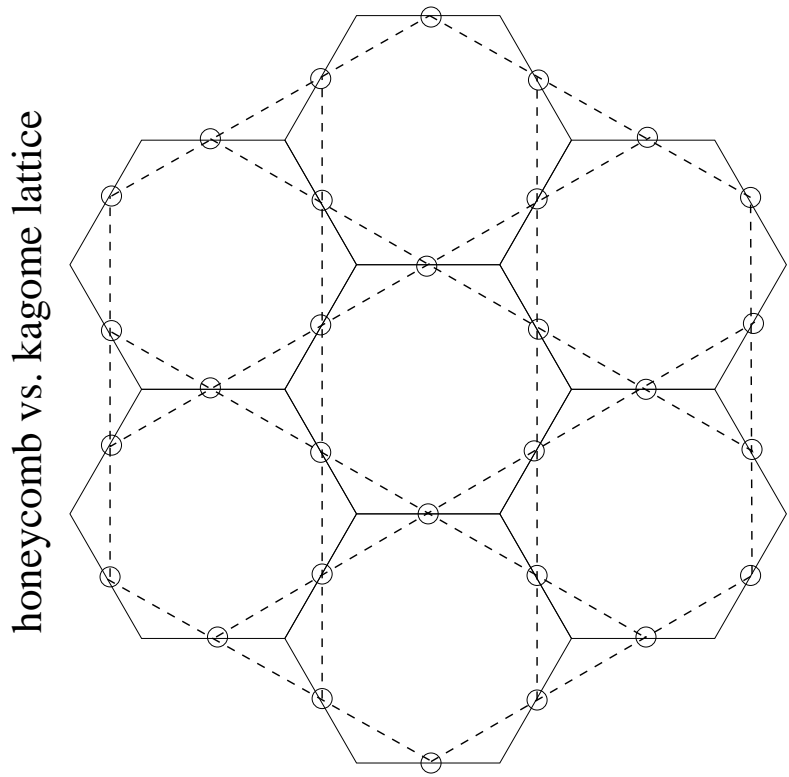

Figure 9: Construction of the kagome lattice as the line graph of the honeycomb lattice.

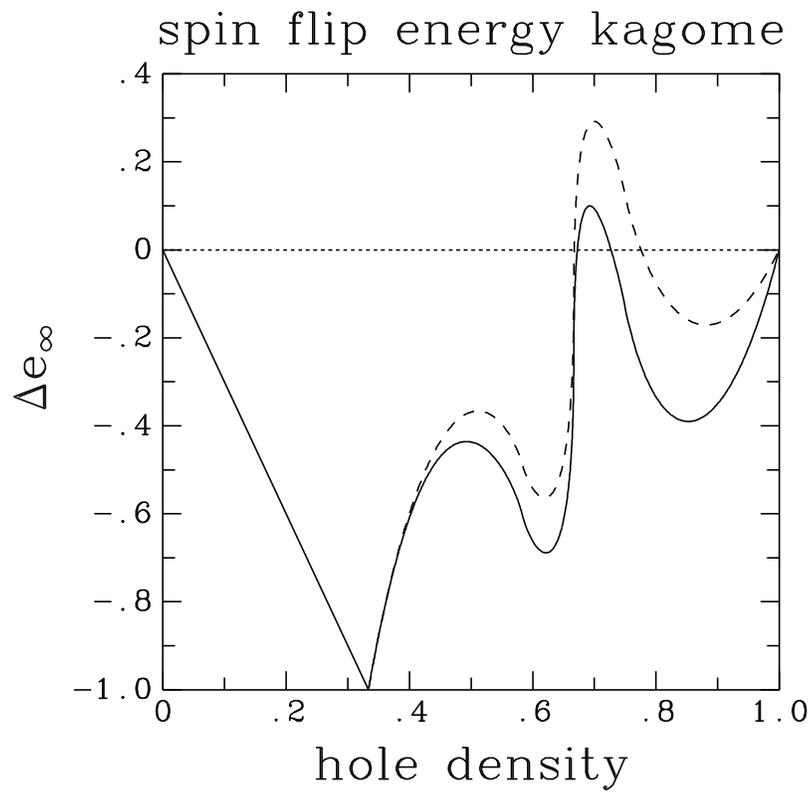

Figure 10: Spin flip energy $\Delta e_\infty$ on the kagome lattice at infinite $U$ for the SKA Gutzwiller ansatz (dashed) and the Basile-Elser ansatz with nearest neighbour hopping (full line), for less than half filling ($n < 1$). The regime $0 \leq \delta \leq 1/3$ corresponds to the flat band at the upper band edge.

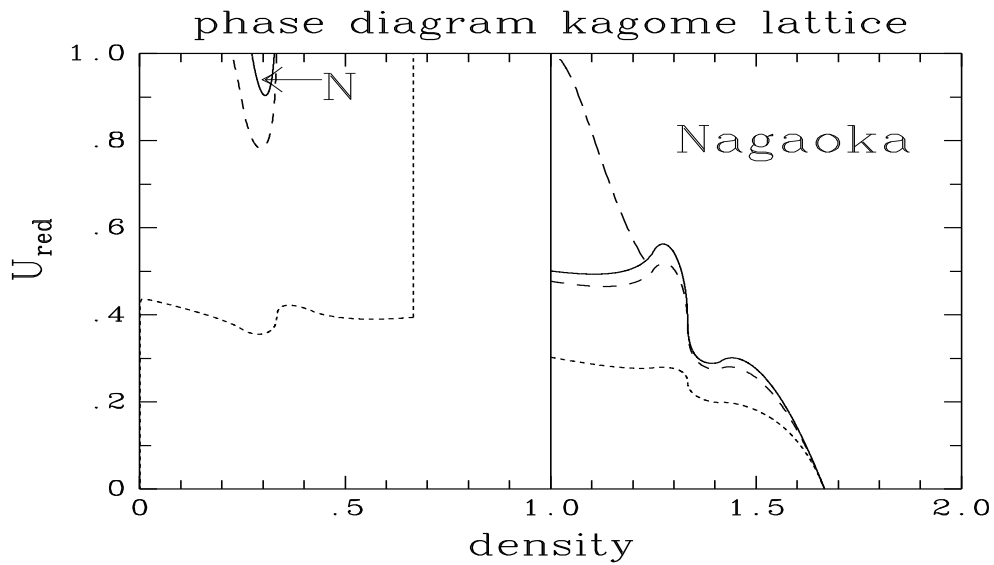

Figure 11: Phase diagram for the kagome lattice: Nagaoka instability lines obtained with the Hartree single spin flip ansatz (dotted), the SKA Gutzwiller ansatz (short dashed) and the Basile-Elser ansatz with nearest neighbour hopping (full line). The global instability against phase separation (long dashed) is also shown. The regimes $2/3 \leq n \leq 1$ and $5/3 \leq n \leq 2$ correspond to the flat bands. $U_{BR} = 13.80|t|$